# Experimental creation and annihilation of nonvolatile magnetic skyrmions using voltage control of magnetic anisotropy without an external magnetic field

Dhritiman Bhattacharya[1], Seyed Armin Razavi[2], Hao Wu[2], Bingqian Dai[2], Kang L. Wang[2*] and Jayasimha Atulasimha[1, 3*]

[1]*Dept. of Mechanical and Nuclear Engineering, Virginia Commonwealth University, Richmond, VA 23284, USA*
[2]*Dept. of Electrical and Computer Engineering, University of California, Los Angeles, CA 90095, USA*
[3]*Dept. of Electrical and Computer Engineering, Virginia Commonwealth University, Richmond, VA 23284, USA*
**Corresponding authors:* wang@seas.ucla.edu *and jatulasimha@vcu.edu*

**Abstract:** In this work, we utilize voltage controlled magnetic anisotropy (VCMA) to manipulate magnetic skyrmions that are fixed in space. Memory devices based on this strategy can potentially be of smaller footprint and better energy efficiency than current-controlled motion-based skyrmionic devices. To demonstrate VCMA induced manipulation of skyrmions, we fabricate antiferromagnet/ferromagnet/oxide heterostructure films where skyrmions can be stabilized without any external magnetic field due to the presence of exchange bias. These isolated skyrmions were annihilated by applying a voltage pulse that increased PMA. On the other hand, decreasing PMA promoted formation of more skyrmions. Furthermore, skyrmions can be created from chiral domains by increasing PMA of the system. To corroborate our experimental observations, we performed micromagnetic simulation. The proposed method could potentially lead to novel skyrmion-based memory devices.

Following the first experimental observation in bulk MnSi [1], magnetic skyrmion related research has experienced a remarkable growth over the last decade. Due to its topological spin structure, magnetic skyrmions have significantly lower depinning current compared to domain walls [2]. This has motivated extensive studies on skyrmion motion based implementation of racetrack memory [3] as well as logic devices [4]. Several experimental studies were performed to understand the skyrmion dynamics that could lead to successful implementation of such devices [5–9]. However, these devices may necessitate larger footprint to accommodate skyrmion motion. In this paper, we demonstrate manipulation of static (not moving) isolated magnetic skyrmions using Voltage Control of Magnetic Anisotropy (VCMA). This strategy has potential to be energy efficient as the manipulation of the skyrmionic state is achieved via VCMA through the application of an electric field and without needing a current or magnetic field.

As spin spiral structures such as magnetic skyrmions are stabilized due to a balance of Perpendicular Magnetic Anisotropy (PMA) and Dzyaloshinskii-Moriya Interaction (DMI) [10,11], one can expect to achieve a magnetic state change via tuning the balance between PMA and DMI. Such tuning can be achieved by adjusting the PMA of the system employing Voltage Control of Magnetic Anisotropy (VCMA) [12,13]. This occurs as application of an electric field modifies the electron density at ferromagnet/oxide interface, which consequently changes the PMA [12]. We have previously shown skyrmion creation, annihilation, reversal as well as skyrmion mediated ferromagnetic reversal in confined geometries by modulating PMA [14–16] through micromagnetic simulations. These results suggest that skyrmion and skyrmion mediated ferromagnetic reversal can implement memory devices where the switching error is extremely robust to defects and thermal noise [14]. Recently, there are also some reports that experimentally investigate electric field induced control of skyrmions. For example, Scanning Probe Microscope (SPM) tip was used to apply an electrical field to Fe monolayer [17] to create skyrmions at very low temperatures. In another work, skyrmion manipulation at room temperature occurs primarily as a result of $M_s$ change due to change in Curie temperature caused by the application of an electric field [18] using an ITO electrode. In that work, the number of skyrmions vary as a function of the voltage/electric field applied and are hence volatile (change when the applied electric field is removed). Finally, room temperature creation and movement of skyrmion bubbles were reported when domain walls are moved in an electric field induced magnetic anisotropy gradient due to geometric effects [19]. In this work, we report experiments that demonstrate voltage control of skyrmions that are fixed in space. These skyrmions are stabilized without any external field and can be annihilated by applying a negative voltage pulse while application of an opposite polarity (positive) voltage pulse can be used to create skyrmions. Additionally, we show that skyrmions can be created from chiral domains. Furthermore, these skyrmion/ferromagnetic states, once created, do not change when the applied electric field is removed and are hence nonvolatile. Finally, we perform micromagnetic simulations to corroborate these findings. We therefore show that electric field induced nonvolatile creation and annihilation of skyrmions is possible without using any external magnetic field. This mechanism could potentially pave the way for manipulation of skyrmion states in confined geometries such as a free layer of a Magnetic Tunnel Junction (MTJ) to achieve non-volatile skyrmion based magnetic memory [14].

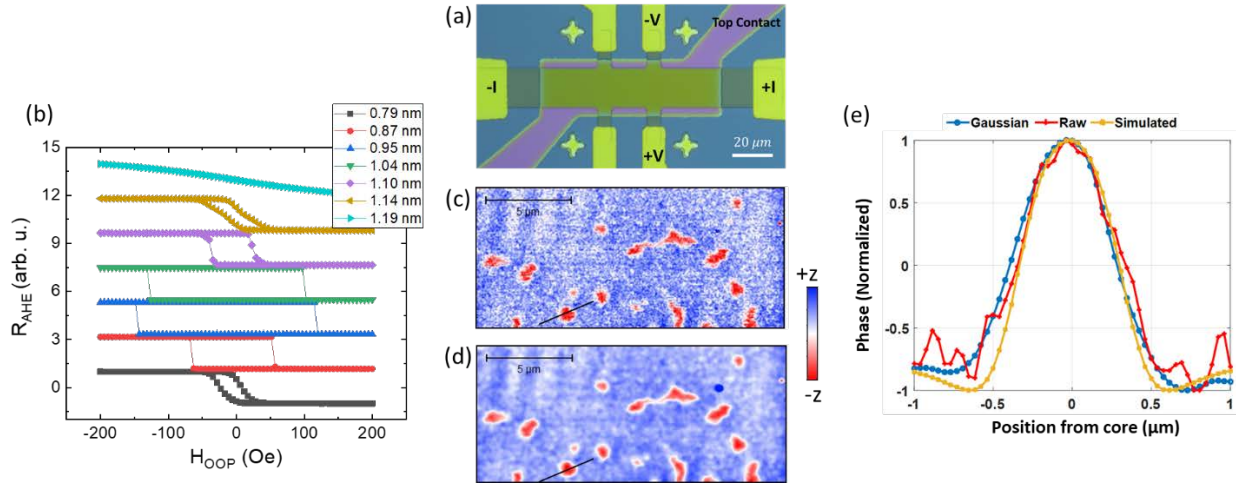

Figure 1. (a) Device structure: The two current contacts can be used for applying current through the stack and the voltage contacts can be used for measuring transverse voltage. There are two similar sets of voltage contacts. These are used for Hall measurements. For VCMA, a voltage pulse was applied between one of the two top gate contacts, and any one of the other current and voltage contacts. (b) Anomalous Hall measurements for different CoFeB thicknesses, where $H_{OOP}$ denotes out-of-plane magnetic field, (c) Raw magnetic force microscopy image showing zero field skyrmions, for CoFeB thickness = 1.126 nm (d) Filtered image for better visualization, (e) Skyrmion profile: Raw, filtered and simulated obtained from the line shown in (c) & (d)

The structure used in our experiments is Ta (2) / IrMn (5) / CoFeB (0.52-1.21) / MgO (2.5) / $Al_2O_3$ (35)/ ITO where the numbers represent the thicknesses in nm. Details of fabrication procedure can be found in the methods section. The interfaces of CoFeB with the antiferromagnetic IrMn and the oxide layer give rise to DMI and PMA. Similar to conventional VCMA induced switching, ferromagnetic/oxide interface is used to achieve the necessary PMA as well as enable VCMA. Additionally, in this heterostructure, PMA depends on the antiferromagnetic layer thickness. On the other hand DMI may exist in both these interfaces [20,21], which could lead to additive DMI. Across the wafer, an array of several hall bars (130×20 $\mu m$) was fabricated (Fig. 1(a)). The thickness of the CoFeB layer was varied across the wafer roughly between 0.52 nm to 1.21 nm. The magnetic properties of the devices, especially the PMA and the DMI, are expected to vary with the thickness of the CoFeB layer. To verify this, anomalous hall resistance due to Anomalous Hall Effect (AHE) was measured to estimate the magnetization component perpendicular to the film upon application of a perpendicular external magnetic field. This is shown in Fig. 1(b) where the hysteresis loops obtained show the expected trend. For example, devices in the range of 0.87 nm - 1.1 nm CoFeB layer exhibit higher perpendicular anisotropy and abrupt switching, while devices with thickness on either side of this range showed lower perpendicular anisotropy and gradual transition during reversal. Another important observation is the presence of exchange bias field in all the devices (5-20 Oe) that emerges from the ferromagnet/antiferromagnet interface. While presence of PMA and DMI are adequate, an external bias magnetic field is generally required to stabilize skyrmions during the experiments. However, in our structure, the readily available interfacial exchange bias field eliminates this requirement thus allowing stabilization of skyrmions at zero externally applied magnetic bias field [20].

Next, the thickness dependence of magnetic states was directly imaged using Magnetic Force Microscopy (MFM). In devices with high perpendicular anisotropy, the magnetization orientation was found to be completely out of plane. In devices with increased thickness (that showed gradual switching), we observed stripe domains and skyrmions. The magnetic states as a function of thickness characterized using MFM is shown in supplementary Fig. S1. In Fig. 1 (c) we show raw MFM phase image of isolated skyrmions and chiral domains at one such thickness, t = 1.126 nm. Across the device, the diameter of the skyrmions vary widely. The raw image was further processed using a Gaussian filter to aid visualization by eliminating noise, which is shown in Fig. 1 (d). All subsequent MFM images are processed using this method. The raw images are shown in the supplementary section Fig. S9. Fig. 1 (e) shows a skyrmion profile that is extracted from the MFM phase image along the line shown in Fig. 1(c, d). The simulated MFM profile of a 475 nm radius skyrmion reasonably matches with the experimental observation. Details of the micromagnetic simulation are discussed later in the paper.

The MFM imaging was further complemented with magneto optical Kerr effect (MOKE) imaging in one such device and domains and skyrmions were clearly observed in the intermediate steps of the switching during a cycle of perpendicular magnetic field as shown in supplementary Video 1. We note that, either of these two imaging technique (MOKE/MFM) cannot resolve the detail of the spin spiral. In other words, these technics are not sufficient to confirm the detailed topological nature of the magnetic objects observed and hence the experimental data is interpreted in conjunction with the simulations that can show the details of the spin spiral. The DMI present in such a structure can be around 100 μJ/m$^2$ [21] and was proven to be sufficient to stabilize both isolated skyrmions and chiral domains as confirmed by our micromagnetic simulation. Further, in a similar film stack, current induced motion of these magnetic objects exhibited skyrmion Hall Effect, which is a clear indication of the topological nature of these magnetic objects (see supplementary Fig. S2).

**Manipulation of skyrmions with electric field alone in the absence of an external magnetic field:** Thereafter, we probe the effect of application of an electric field in the devices where skyrmions and domains were observed at zero magnetic field. A voltage pulse is applied between one of the two top gate contacts, and any one of the other current and voltage contacts. Current and voltage contacts are all attached

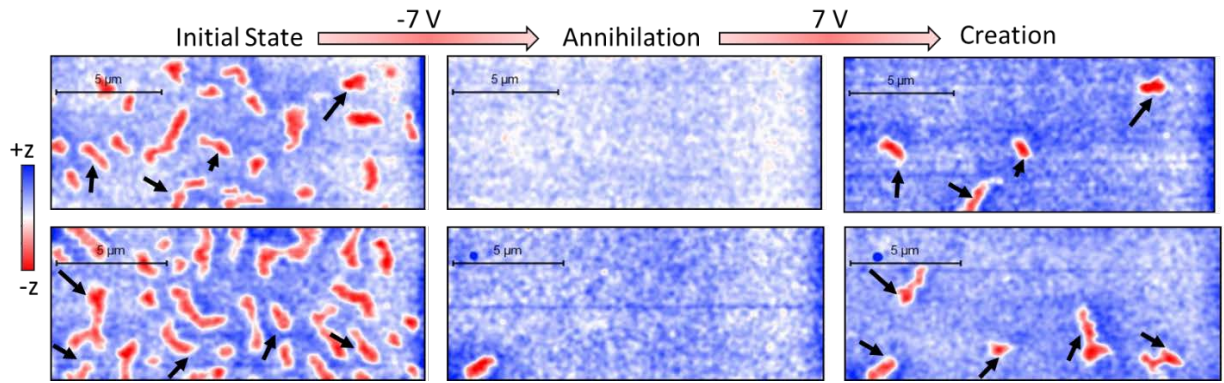

*Figure 2. MFM images obtained before and after application of electric field. Left column: magnetization state before application of any electric field. Middle Column: Magnetization state obtained after applying a negative voltage pulse that increased PMA. Right Column: Magnetization state obtained after applying a positive voltage pulse that decreased PMA.*

to the Hall bar, which is metallic and conductive. In this way, the electric field is dropped over the MgO and Al$_2$O$_3$. Consequently, the electron density at the ferromagnet/oxide interface is modulated, leading to a modulation in the PMA of the system. The oxide barrier breakdown voltage is around 8 V. In most of the cases, we applied ±7 V to our structure (unless otherwise mentioned) by using a Keithley 2636B source meter and Signatone probe station, which corresponds to an electric field of $E_{MgO} = \frac{V}{t_{MgO} + \frac{\epsilon_{MgO}}{\epsilon_{AlOx}} t_{AlOx}} = 0.157$

V/nm, using $\frac{\epsilon_{MgO}}{\epsilon_{AlOx}} = \frac{9}{7.5}$. In our system, application of a positive (negative) electric field leads to decrease (increase) of the PMA. Before applying this electric field, we imaged the initial magnetization state of the device as shown in the left column of Fig. 2. The initial states consist of skyrmions and stripe domains. Next, we applied -7 V between the top and the bottom electrode for 1-2 seconds to increase the PMA of the system. The voltage pulse was withdrawn and the transformation of the magnetic state due to application of this voltage pulse was imaged in the absence of an applied electric field. We observed that the skyrmions (and the stripe domains) were annihilated (middle column of Fig. 2) and the magnetization of the system reoriented in the +z direction as evidenced by the MFM image. Subsequently, we applied an opposite polarity voltage pulse (i.e. +7V) in a similar manner. Due to this, the PMA decreases and DMI prevails over PMA. This is expected to be a favorable condition for formation of spin spiral states. Indeed, some skyrmions and chiral domains reappear as can be seen in the MFM images shown in the right column of Fig. 2. We note that all imaging was performed at zero external magnetic field and zero applied electric field. Therefore, these creation and annihilation processes were nonvolatile and were achieved without the assistance of any external bias magnetic field. Therefore, skyrmionic state and saturated out of plane ferromagnetic state both emerged as stable states of the system and the applied electric field can result in the switching between these states.

Additionally, we observed incomplete annihilation where starting from an initial state with mostly chiral domains and some skyrmions, a negative voltage pulse could annihilate some of the skyrmions while creating many from the stripe domains as shown in Fig 3. Therefore, transformation from chiral domain to skyrmions is also achievable using VCMA. However, unlike the previous case, this transformation was found to be irreversible. In other words, application of a positive pulse could not recreate more skyrmions or transform the skyrmions back to original chiral domains. In the case of incomplete annihilation, many of the stripe domains turn into skyrmions where they get pinned more strongly. Therefore, subsequent application of a positive pulse (i.e. reduction of PMA) did not affect it substantially. Transformation from chiral domain to skyrmions were previously observed experimentally by using a current pulse [6,7]. Theoretically, chopping skyrmions from chiral domain was shown using strain [22]. Here, we demonstrate the feasibility of such transformation by employing VCMA.

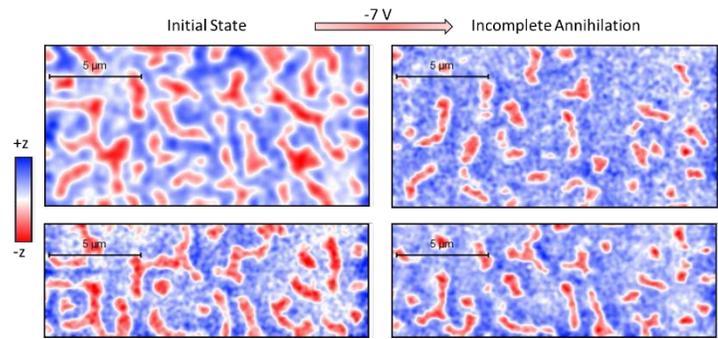

Figure 3. Incomplete annihilation that shows chiral domain to skyrmion transformation.

All these different observations can be satisfactorily explained if we consider pinning sites in the thin film stack. The CoFeB and the IrMn films are polycrystalline in nature, which could lead to intergranular variation of PMA, DMI and exchange bias. Pinning sites could also exist due to inhomogeneity in film thickness and presence of defects etc. We observed the size of the skyrmions vary widely. Additionally, these skyrmions are not perfectly circular. These indicate the existence of inhomogeneity. Due to corresponding non-uniformity of magnetic parameters across the film, the chiral domains and the skyrmions have a propensity to occupy the same location. In successive magnetization cycles, the stripe domains and skyrmions were observed to emerge roughly at the same position (supplementary Fig. S3). Similarly, skyrmions created by VCMA appear at the same initial location that they occupied before annihilation as indicated by the arrows in the top panel of Fig. 2. The same skyrmions were also present at the same location in all the magnetization cycles (see supplementary Fig. S3).

Finally, as shown in Fig. 2, we observed lesser number of skyrmions were created by VCMA compared to the initial state (prior to annihilation). It was previously shown that due to non-uniformity, more skyrmions can be created from a stripe domain state than from a single domain saturated state [20]. In our study, the initial state was prepared by first saturating the device under a large magnetic field and then subsequently reducing the field to zero. While the field is reduced, stripe domains emerged and many skyrmions were created from this stripe domains. Upon complete annihilation of these skyrmions and strip domains with a negative electric field pulse, we are left with a saturated magnetic state. Due to subsequent application of a positive electric field pulse, skyrmions are now created from an out-of-plane saturated state (after the first annihilation cycle). Therefore, fewer skyrmions were created.

**Simulations for manipulation of skyrmions with electric field:** Next we perform micromagnetic simulation to show the voltage controlled creation and annihilation process of skyrmions. Details of the simulation procedure and parameters used can be found in the methods section. Although due to inhomogeneity, PMA, DMI and exchange field vary across the device, the effect is simulated by varying only one of the parameters for simplicity namely perpendicular anisotropy. To illustrate skyrmion creation and annihilation, we define regions of low perpendicular anisotropy to incorporate pinning sites. The lower the anisotropy, the stronger the pinning site. In a 4 μm × 2 μm rectangular geometry, three circular pinning sites of varying strength and 500 nm diameter were defined (Fig. 4, left column). Initially skyrmions were stabilized at each of these pinning sites (Fig. 4, second column). The effect of voltage application is primarily limited to changing the perpendicular magnetic anisotropy of the system. In other words, applied voltage does not alter the saturation magnetization, exchange bias etc. as discussed in supplementary section 9. There are recent reports of DMI modulation by applying an electric field [23]. However, these changes in DMI are relatively small and do not affect the magnetization dynamics significantly [14]. Therefore, we only consider the change in PMA. VCMA coefficient in a similar heterostructure was measured to be 38.2 fJ/Vm (Supplementary Fig. S4). Therefore, electric field of 0.157 V/nm can cause $6\times10^3$ J/m$^3$ change in anisotropy energy density (i.e. interfacial PMA ($k_i$)/thickness). We considered a change of $5\times10^3$ J/m$^3$ in anisotropy density which resulted in annihilation of all but the most strongly pinned skyrmion (Fig. 4, third column). However, from this annihilated state, only one new skyrmion formed when PMA is decreased. This skyrmion forms at the second strongest pinning site. Similarly, assuming other configurations of PMA distribution it can be shown that stripe domains can be transformed to skyrmions (Fig.4, Bottom panel). Although simplistic, these simulations are representative of the experimentally observed skyrmion creation and annihilation process.

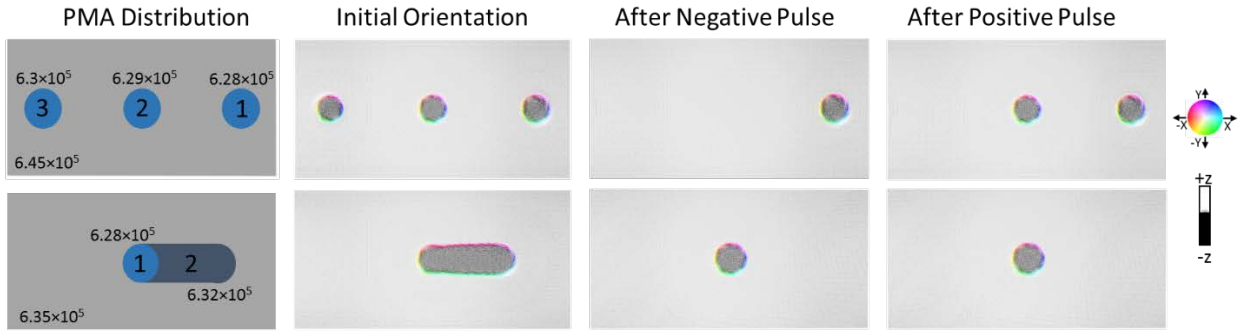

*Figure 4. Top panel: Skyrmion annihilation and creation by VCMA. The numbers show the PMA in J/m$^3$. Pinning strength: 1>2>3. Bottom Panel: Creation of skyrmion from a chiral domain.*

In summary, we have demonstrated VCMA induced control of magnetic skyrmions. Particularly, we have shown stabilization of skyrmions without applying any external magnetic field that could be annihilated and recreated by applying voltages of opposite polarity. Additionally, skyrmion creation from chiral domain state was also shown. This control was nonvolatile and dependent on the pinning sites across the device. These observations were corroborated by our micromagnetic analysis. Such electric field control of magnetic skyrmions could pave the pathway of energy efficient high-density magnetic memory devices.

## Methods

**Fabrication/MOKE/Transport:**

The layers consisting of Ta (2) / IrMn (5) / CoFeB (0.52-1.21) / MgO (2.5) / Al$_2$O$_3$ (5) were grown on Si/SiO$_2$ substrates by dc and rf magnetron sputtering at room temperature, where the numbers represent the thicknesses in nm. The CoFeB layer has a wedge shape with continuously changing thickness. The samples where then patterned into an array of Hall bar devices using standard photolithography techniques. A 30 nm Al$_2$O$_3$ gate oxide was deposited using atomic layer deposition (ALD), and ITO layers were fabricated as a top gate electrode. The samples were then annealed at 150$^o$C for 30 minutes under an out-of-plane magnetic field of 6 kOe to introduce the exchange bias and enhance the perpendicular anisotropy. The dimensions of the Hall bars are 20 $\mu m$ × 130 $\mu m$. All electrical and optical measurements were done at room temperature using Keithley 6221 current source, Keithley 2182A nanovoltmeter, Stanford Research Systems SR830 lock-in amplifier, and HeNe laser with 632.8 nm of wavelength. The external magnetic field is provided by an electromagnet driven by a Kepco power supply.

**Magnetic Force Microscopy:**

We obtained MFM image at room temperature and atmospheric pressure with Bruker Dimension Icon AFM system. We used Bruker MESP-LM low-moment probes to minimize tip-induced magnetization reorientation. To confirm there is no tip induced effects, we scanned the same area twice (scanning up and down). These two scans produced similar images (see supplementary section, Fig. S5). Nominal cantilever frequency, lift height and scan rate were respectively 75 kHz, 40 nm and 0.2 Hz.

**Micromagnetic Simulation:**

Micromagnetic simulation software-Mumax3 [24] was used to perform the simulations where the magnetization dynamics is simulated using the Landau-Lifshitz-Gilbert (LLG) equation:

$$\frac{\partial \vec{m}}{\partial t} = \vec{\tau} = \left(\frac{-\gamma}{1+\alpha^2}\right)\left((\vec{m} \times \vec{H}_{eff}) + \alpha\left(\vec{m} \times (\vec{m} \times \vec{H}_{eff})\right)\right)$$

Here $\vec{m}$ is the reduced magnetization ($\vec{M}/M_{sat}$), M$_{sat}$ is the saturation magnetization, $\gamma$ is the gyromagnetic ratio and $\alpha$ is the Gilbert damping coefficient.

The quantity $H_{eff}$ is the effective magnetic field, which includes the effective field due to demagnetization energy, exchange bias, Heisenberg exchange coupling and DMI interaction, and perpendicular magnetic anisotropy (PMA).

$$\vec{H}_{eff} = \vec{H}_{demag} + \vec{H}_{exch\_bias} + \vec{H}_{exchange} + \vec{H}_{anis}$$

H$_{anis}$ is the effective field due to the perpendicular anisotropy

$$\vec{H}_{anis} = \frac{2K_{u1}}{\mu_0 M_s}(\vec{u} \cdot \vec{m})\vec{u} + \frac{4K_{u2}}{\mu_0 M_s}(\vec{u} \cdot \vec{m})^3$$

where, $K_{u1}$ and $K_{u2}$ are first and second order uniaxial anisotropy constants and $\vec{u}$ is the unit vector in the direction of the anisotropy (i.e. perpendicular anisotropy in this case). VCMA effectively modulates the anisotropy energy density, which is given by ΔPMA =$\xi E$. Here $\xi$ and E are respectively the coefficient of electric field control of magnetic anisotropy and the applied electric field. The resultant change in uniaxial anisotropy due to VCMA is incorporated by modulating $K_{u1}$ while keeping $K_{u2} = 0$.

Table I

| Parameters | Value |
|---|---|
| Saturation Magnetization ($M_s$) | $1 \times 10^6$ A/m |
| Exchange Constant (A) [20] | $1.1 \times 10^{-11}$ J/m |
| Perpendicular Anisotropy Constant ($K_{u1}$) | $6.35 \times 10^5$ J/m$^3$ |
| DMI Parameter (D) [21] | 100 µJ/m$^2$ |
| Gilbert Damping ($\alpha$) | 0.1 |
| Exchange Bias | 5 Oe |

**Supporting Information:** Thickness dependent evolution of magnetic states, Skyrmion Hall effect, emergence of skyrmion/stripe domains at same location in different magnetization cycle, VCMA co-efficient, Up and Down MFM Scan, other examples (more cases), exchange bias and $M_s$ dependence on gate voltage and raw MFM images.

**Acknowledgment:** D.B and J.A are supported in part by the NSF CAREER grant CCF-1253370 and NSF ECCS 1609303, VCU Quest Commercialization Grant and Virginia Microelectronics Seed Grant. S. A. R., H. W., B. D., and K. L. W. are supported by the National Science Foundation (NSF) ECCS 1611570 and NSF Nanosystems Engineering Research Center for Translational Applications of Nanoscale Multiferroic Systems (TANMS). The authors at UCLA are also supported by the Spins and Heat in Nanoscale Electronic Systems (SHINES), an Energy Frontier Research Center funded by the U.S. Department of Energy (DOE), Office of Science, Basic Energy Sciences (BES) under Award No. SC0012670. The authors at UCLA were also partially sponsored by the Army Research Office under Grant Number W911NF-16-1-0472.

# Supplementary Information

# Experimental creation and annihilation of nonvolatile magnetic skyrmions using voltage control of magnetic anisotropy without an external magnetic field

Dhritiman Bhattacharya[1], Seyed Armin Razavi[2], Hao Wu[2], Bingqian Dai[2], Kang L. Wang[2*] and Jayasimha Atulasimha[1, 3*]

[1]Dept. of Mechanical and Nuclear Engineering, Virginia Commonwealth University, Richmond, VA 23284, USA
[2]Dept. of Electrical and Computer Engineering, University of California, Los Angeles, CA 90095, USA
[3]Dept. of Electrical and Computer Engineering, Virginia Commonwealth University, Richmond, VA 23284, USA
*Corresponding authors: wang@seas.ucla.edu and jatulasimha@vcu.edu

1. **Thickness dependent magnetic state:** The magnetization state in the studied device depends on the thickness of the ferromagnetic CoFeB layer as magnetic parameters such as PMA and DMI vary with thickness. This is shown in Fig. S1 at zero external magnetic field using Magnetic Force Microscopy (MFM) images at different thicknesses (t= 1.092-1.143 nm).

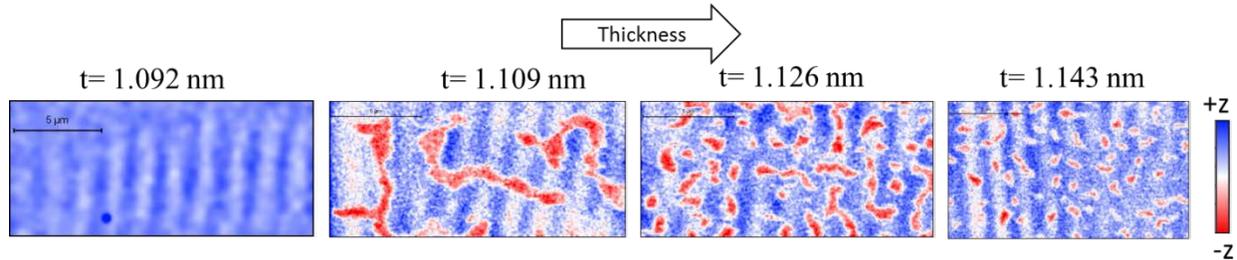

*Figure S1. Evolution of magnetic state with thickness.*

2. **Skyrmion Hall effect:** We utilize spin-orbit torques from IrMn (with a thickness of 4 nm) to drive the skyrmion motion [1]. It has been shown that IrMn has a sizable spin Hall angle [2], allowing for a relatively large damping-like spin-orbit torque. First, we create skyrmions by scanning the external magnetic field, and then by utilizing electrical current in this system, we displace skyrmions by pulses with an amplitude of 15 mA and a duration of 6 ms. As seen in Fig. S2, the direction of the current-driven skyrmion motion is not along the current direction and has a transverse component. This phenomenon is the so-called skyrmion Hall effect [3], which is a signature of magnetic skyrmions. These results suggest that the bubble-like particles that we observe have a topological nature and are indeed skyrmions.

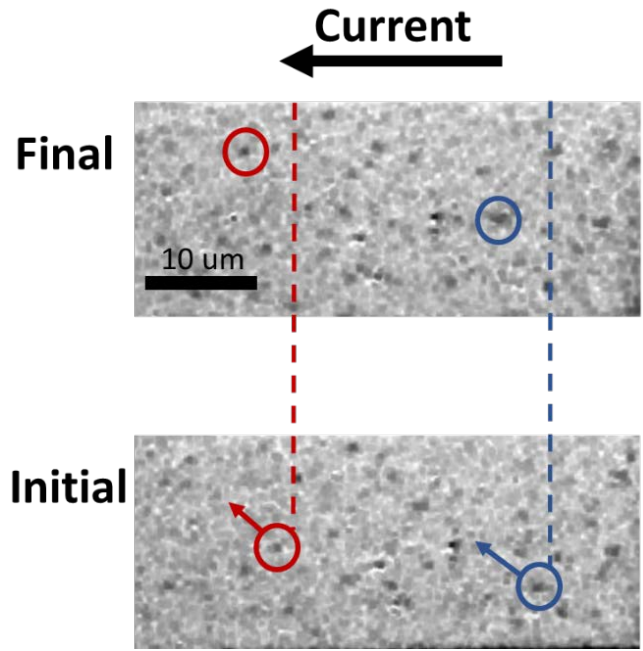

*Figure S2. Current-driven skyrmion motion imaged using magneto-optical Kerr effect (MOKE) microscopy. Red and blue circles indicate the initial and final positions of two tracked skyrmions, before and after applying current pulse. The transverse motion of skyrmions is due to the skyrmion hall effect.*

### 3. Emergence of skyrmion/stripe domains at the same location in different magnetization cycle:

We magnetized our sample several times by first applying a large saturation magnetic field and subsequently removing the field to observe the equilibrium magnetic states. We found skyrmions and stripe domains have a propensity to nucleate/form at the same location as shown in Fig. S3. This demonstrates that there is inhomogeneity (i.e. nucleation/pinning sites) across the device and due to that, skyrmions/chiral domain are created preferentially in these regions of the film. Moreover, the skyrmions that were created by electric field application were also present in the different magnetization cycles. This perhaps indicates that these are the most favorable sites for skyrmion/domain nucleation among all the possible nucleation/pinning sites. These positions are shown with arrows.

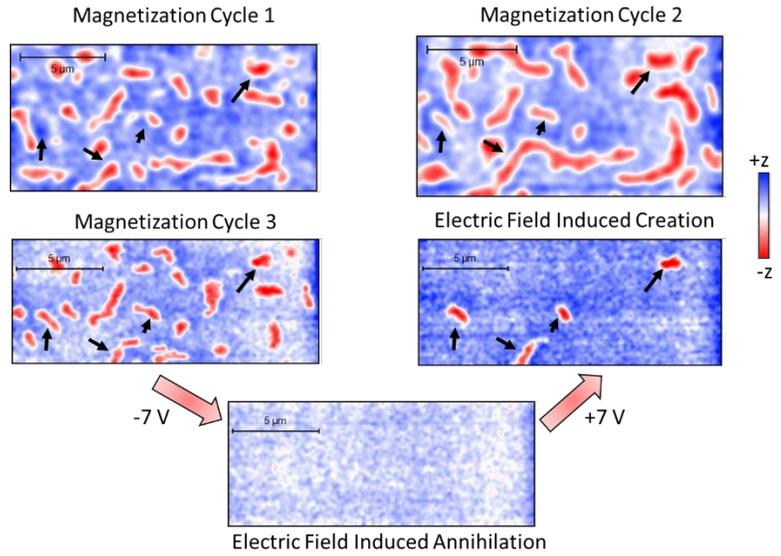

*Figure S3. Skyrmions/stripe domains at almost exact position in different magnetization cycle and after electric field induced creation.*

### 4. VCMA co-efficient:
We characterize VCMA coefficient in our perpendicularly magnetized samples by measuring anomalous Hall resistance as a function of an in-plane magnetic film in our Hall bar structures with top gates. We can then calculate the interfacial anisotropy using the transport data and magnetization saturation obtained from SQUID measurement (~1050 emu/cc) [4]. The results are shown in Fig. S4, where interfacial anisotropy is plotted as a function of applied electric field. VCMA coefficient ($\xi$) is defined as the slope of this plot.

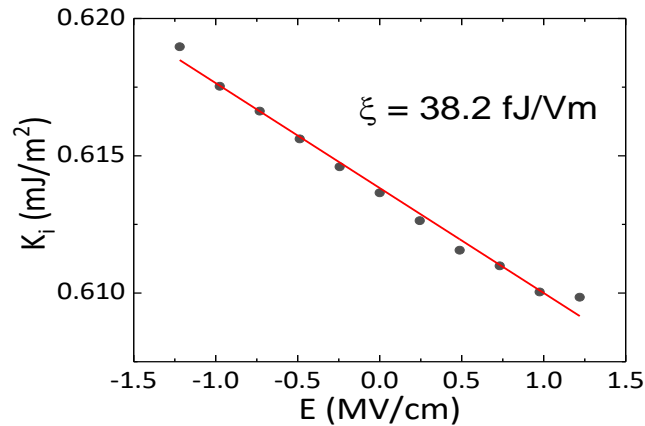

*Figure S4. Interfacial anisotropy ($K_i$) as a function of the applied electric field (E) in the sample with CoFeB thickness of 1.06 nm. The slope of this plot is the VCMA coefficient ($\xi$).*

5. **Up and Down Scan:** To confirm there is no tip induced effects, we scanned the same area twice (scanning up and down). These two scans produced similar images as shown in Fig. S5 indicating that tip induced changes to the magnetic state of the films in minimal.

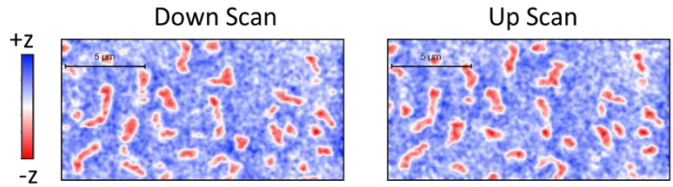

*Figure S5. MFM image of the same location scanning up and down*

6. **Other Examples:** We performed similar study staring from an initial state that had opposite polarity (core-up) skyrmions as shown in Fig. S6. These skyrmions were found to be larger in size than the core-down skyrmions as the exchange bias field is pointing in the same direction (+z direction or upwards). Application of a voltage pulse that increased PMA was utilized to annihilate these skyrmions/domains and the magnetization saturates in the -z direction. However, skyrmions could not be recreated by applying a positive voltage pulse possibly due to pinning effect as explained in the main text.

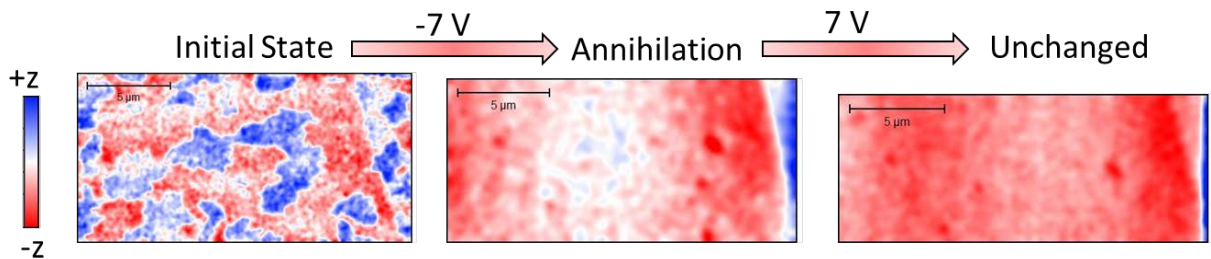

*Figure S6. Other examples of voltage control of skyrmions*

Some other examples of incomplete annihilation are shown in Fig. S7 where some skyrmions were annihilated while chiral domains transformed into skyrmions due to application of a negative voltage pulse.

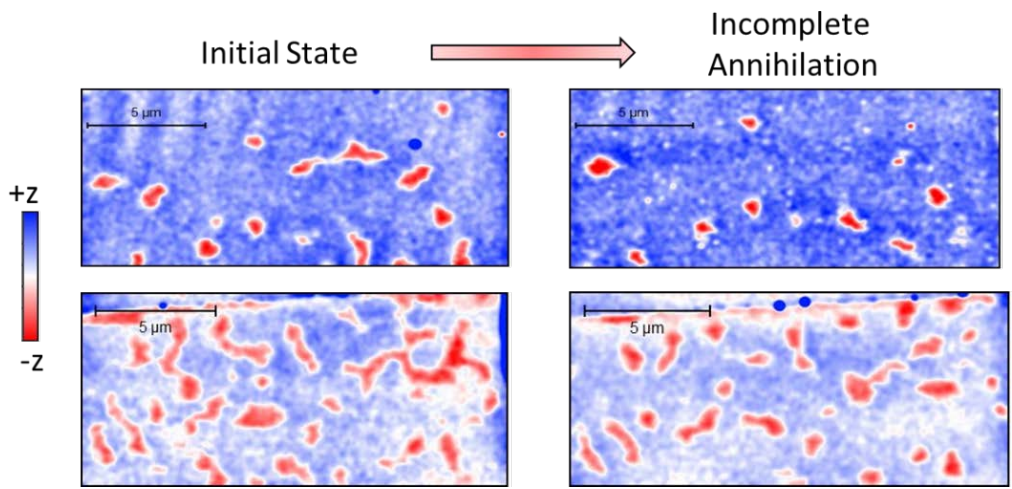

*Figure S7. Other examples of incomplete annihilation due to a negative voltage pulse*

**7. Exchange bias and $M_s$ dependence on gate voltage:** In order to investigate the effect of the applied voltage on the exchange bias and magnetization saturation in our sample, we did anomalous Hall effect measurements under different applied gate voltages as shown in Fig. S8. This measurement is done on the sample with CoFeB thickness of around 1.08 nm. We observe that the exchange bias under different bias voltages is almost the same and the negligible changes might be due to the change in interfacial properties. Furthermore, the difference in anomalous Hall resistance between the up/down states is independent of the applied voltage, which suggests that the magnetization saturation ($M_s$) does not change.

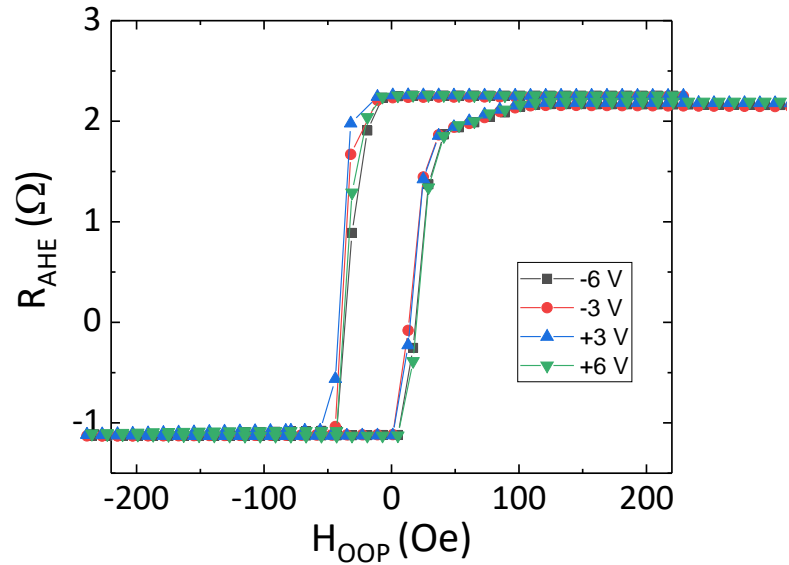

*Figure S8. Anomalous Hall (AHE) measurement on sample with CoFeB thickness of 1.08 nm and under different gate voltages. The exchange bias and the difference in AHE resistance between the two states is almost the same for all samples.*

7. **Raw Images:** Many of the magnetic force microscopy (MFM) images in the main paper and some in the supplement have been presented after performing some image processing. The raw MFM images that correspond to these processed images are presented below.

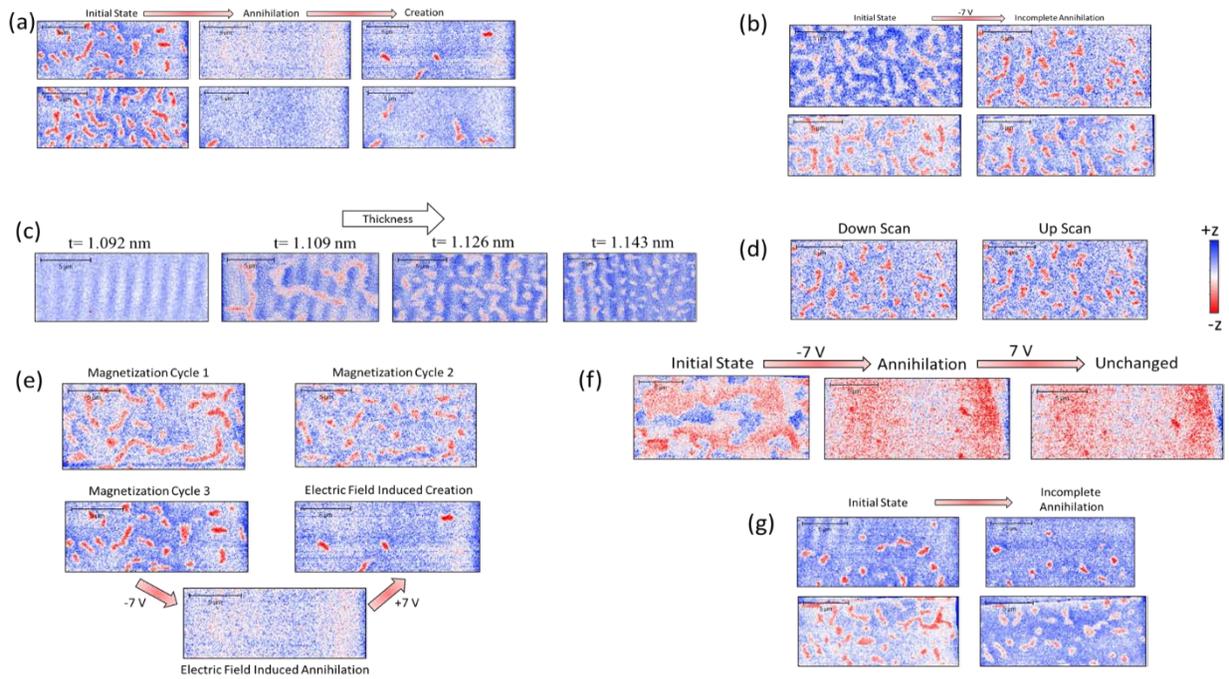

Figure S9. RAW MFM image: (a) Fig. 2, (b) Fig. 3, (c) Fig. S1, (d) Fig. S5, (e) Fig. S3, (f) Fig. S6, (g) Fig. S7